RF compressibility of topological surface and interface states in metal-hBN-Bi$_2$Se$_3$ capacitors

A. Inhofer,[1] T. Wilde,[1,2] J. Duffy,[1,2] M Boukhicha,[1,3] J. Palomo,[1] K. Watanabe,[4] T. Taniguchi,[4] J.M. Berroir,[1] G. Fève,[1] E. Bocquillon,[1] B. Plaçais,[1] B.A. Assaf[1,5]

[1] *Laboratoire de Physique de l'Ecole normale supérieure, ENS, Université PSL, CNRS, Sorbonne Université, Université de Paris, 24 rue Lhomond 75005 Paris, France*

[2] *Department of Chemical Engineering, Northeastern University, 360 Huntington Avenue, Boston MA, 02115, USA*

[3] *Sustainable Energy Technologies Department, Brookhaven National Laboratory, Upton, New York 11973, USA*

[4] *National Institute for Materials Science, 1-1 Namiki, Tsukuba, Ibaraki 305-0044, Japan.*

[5] *Department of Physics, University of Notre Dame, Notre Dame, IN 46556, USA*

**The topological state that emerges at the surface of a topological insulator (TI) and at the TI-substrate interface are studied in metal-hBN-Bi$_2$Se$_3$ capacitors. By measuring the RF admittance of the capacitors versus gate voltage, we extract the compressibility of the Dirac state located at a gated TI surface. We show that even in the presence of an ungated surface that hosts a trivial electron accumulation layer, the other gated surface always exhibits an ambipolar effect in the quantum capacitance. We succeed in determining the velocity of surface Dirac fermions in two devices, one with a passivated surface and the other with a free surface that hosts trivial states. Our results demonstrate the potential of RF quantum capacitance techniques to probe surface states of systems in the presence of a parasitic density-of-states.**

Radio frequency (RF) transport has been widely used to characterize various low-dimensional systems including graphene [1] [2] [3] [4] [5] [6] [7] [8] and to detect and manipulate edge states via interferometric techniques. [9] [10] [11] Such techniques have recently gained importance in the context of the detection of coherent and topological states that are useful for quantum computing. [12] [9] In a non-metallic 2D material, a simple two terminal metal-insulator-2D material capacitor device allows one to directly probe the density-of-states (quantum capacitance) and the channel resistance of the material, without the need for a magnetic field. Previous studies on graphene have successfully used this device geometry to characterize the scattering dynamics of charge carriers and electron-phonon interactions. [1]

RF capacitance measurements in the two terminal device architecture have also been performed on topological insulators. [13] [14] [15] In contrast with graphene, three-dimensional TIs have more complex dielectric properties. Bi$_2$Se$_3$ – a prototypical topological insulator – has a large bulk energy gap (200meV) with two Dirac states bound to each surface. RF capacitance can be used to probe the quantum capacitance ($c_q$), the conductivity of the material and the inter-surface dielectric coupling via the bulk as demonstrated by our recent work. [14] [15] The quantum capacitance $c_q$ is of particular interest since it is proportional to the material's electronic compressibility $\chi$ via $c_q = e^2\chi$. For the Dirac fermions of topological insulators:

$$c_q = \frac{e^2 \Delta \varepsilon_f}{2\pi \hbar^2 v^2} \quad (1)$$

Here $\Delta\varepsilon_f$ is the Fermi energy with respect to the Dirac point, $v$ is the Dirac velocity and $\hbar$ and $e$ have their usual meaning.

In this work, we investigate the quantum capacitance of topological surface states and interface states in metal-insulator-topological insulator capacitor devices (MITI-CAP) having two different architectures. In the first, a top-gate is used to probe the top surface quantum capacitance while in the second a bottom gate probes the quantum capacitance of the substrate-TI interface. We build on the recently demonstrated process that uses physical vapor deposition combined with the dry transfer of a hexagonal boron nitride (h-BN) layer to design and process MITI-CAP devices. [15] [16] A major advantage of such devices is the use of hBN to apply electric fields exceeding 1V/nm and tune the Fermi energy through the surface Dirac point of $Bi_2Se_3$. We successfully achieve this for both the top surface and bottom interface states of $Bi_2Se_3$ and measure the capacitance and channel resistance of MITI-CAPs up to 10GHz in both cases. The devices allow us to measure the velocity of topological Dirac fermions at the gated surface. We compare measurements from two devices with and without an hBN passivation layer. The importance of this passivation layer in preventing surface impurities from yielding an accumulation layer is established. We also demonstrate the viability of RF compressibility measurements to probe surface states even in the presence of the accumulation layer.

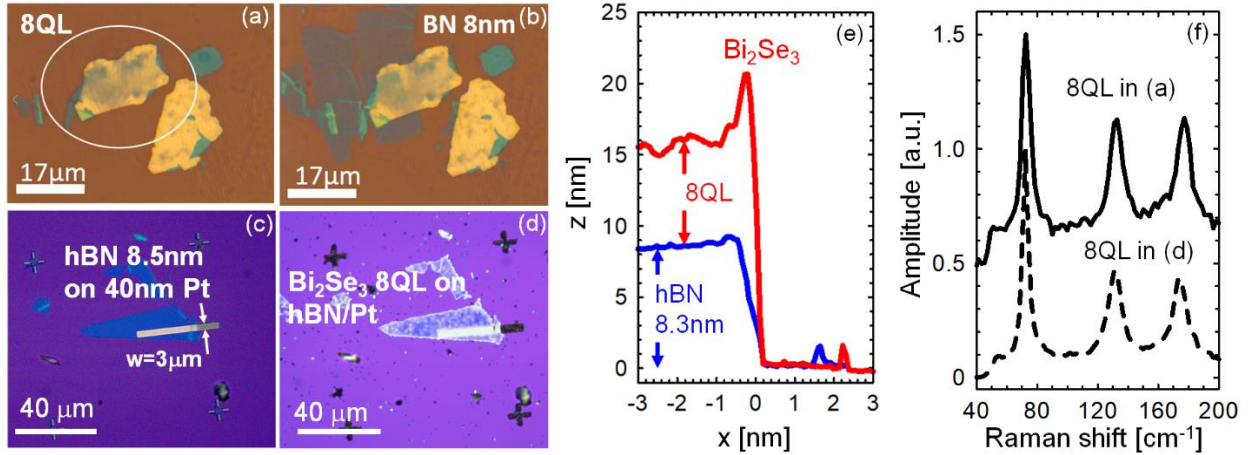

**FIG 1.** (a) $Bi_2Se_3$ flake synthesized by PVD on hBN. (b) Flake shown in (a) with hBN dry-transferred on top. (c) Pt bottom gate with hBN flake dry-transferred on top. (d) $Bi_2Se_3$ synthesized by PVD on the hBN/Pt stack. (e) Atomic force microscope data for the flake shown in (d) before (blue) and after (red) the synthesis of $Bi_2Se_3$. (f) Micro-Raman spectra taken on both samples at room temperatures with a green 514nm laser excitation. (QL: Quintuple layer).

Top-gated MITI-CAPs are processed by first synthesizing $Bi_2Se_3$ on exfoliated hBN using physical vapor deposition (PVD in Fig. 1(a)) and then placing a second hBN layer on top using the well-established dry-transfer technique. [17] [18] We obtain an hBN-$Bi_2Se_3$-hBN heterostructure shown in Fig. 1(b). [16] [15] For the bottom gated MITI-CAP, we first pattern a 40nm Pt gate electrode and then dry transfer an hBN layer on top the Pt gate (Fig. 1(c)). This hBN layer then serves as a substrate for the synthesis of $Bi_2Se_3$ by PVD. A seen in Fig. 1(d), $Bi_2Se_3$ nucleates faster on the Pt/hBN stack than on the $SiO_2$ substrates, leaving a continuous layer deposited to form a Pt-hBN-$Bi_2Se_3$ MITI stack. All samples are characterized using atomic force microscopy and Raman spectroscopy. The top-gated MITI-CAPs were studied in detail in our previous work. [15] An atomic force microscope (AFM) height profile taken on a bottom gated MITI-CAP is shown in Fig. 1(e). By performing AFM on the hBN layer before the growth and on the hBN-$Bi_2Se_3$ after the growth, we can extract the thickness of the hBN layer (8.3±0.3nm) and that of $Bi_2Se_3$ (8 quintuple layers). Note that the typical roughness of our $Bi_2Se_3$ layers is 1 quintuple layer (±1QL) as can be seen in Fig. 1(e). Fig. 1(f) shows Raman spectra taken from both samples shown in Fig. 1. The three characteristic phonon peaks of $Bi_2Se_3$ [19] are observed and confirm the successful synthesis of high-quality material on the hBN surface.

A top-gated MITI-CAP device (CAP1) is finished in a single electron beam lithography step followed by the deposition of a Ti(5nm)/Au(200nm) metal bilayer to form the gate, drain and waveguide. [15] A finished device is shown in Fig. 2(a). The same method is used to deposit the drain and waveguide for the case of the bottom-gated MITI-CAP (CAP2). They are patterned on the free top surface of the $Bi_2Se_3$ by e-beam lithography followed by the deposition of Ti(5nm)/Au(200nm). Lastly, in order to ensure that the $Bi_2Se_3$ is electrically separated from the Pt gate in CAP2, we etch a window in the TI near the gate. A finished device obtained using this hybrid dry-transfer/PVD process in shown in Fig. 2(b) and its insets. The main difference between CAP1 and CAP2 is the location of the gate. In CAP1 illustrated in Fig. 2(c) the gate controls the chemical potential of the top surface and the $Bi_2Se_3$ is sandwiched between hBN layers. In CAP2, shown in Fig. 2(d), the gate controls the bottom surface chemical potential and the top surface is open to air. In both cases a distributed RC model can be used to model the RF response (Fig. 2(e)). The device dimensions are shown in Table 1.

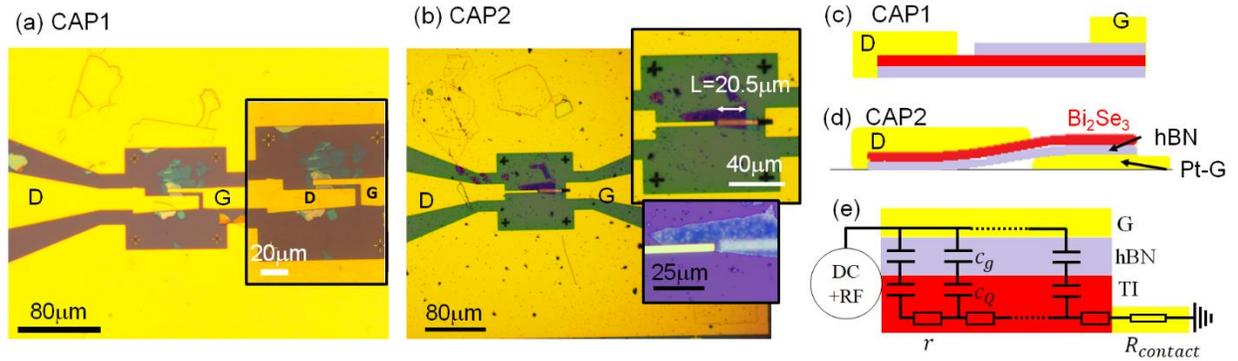

**FIG 2.** Optical microscopy of the two devices investigated in this work (a) CAP1 and (b) CAP2. Sketch of the cross-section of (c) CAP1 and (d) CAP2. D and G label the drain and gate contacts respectively. (e) Circuit diagram of the distributed RC model used to analyze the RF response of both devices.

In what follows, we will focus on the measurement of the quantum capacitance of CAP1 and CAP2 and their comparison. A two-probe setup connected to a network analyzer allows us to determine the S-parameters of the capacitors versus frequency up to 10GHz. A standard short-open-load-through calibration is performed before each measurement at 10K. The experiment is performed with one probe (ground) connected to the drain (D) while RF and DC are simultaneously connected to the gate (G) via a bias-T. The frequency is swept between 70kHz and 10GHz at constant DC gate voltage. The S-parameters are experimentally extracted at different frequencies and converted to admittance $Y$. A de-embedding routine is then performed to disentangle the capacitive and resistive components of a through-line and a dummy device consisting of an identical gate and drain geometry without the sample in between. [20] This procedure is identical to that used in our previous works. [1] [14] [13] It is repeated for different gate voltages (0 to -7V).

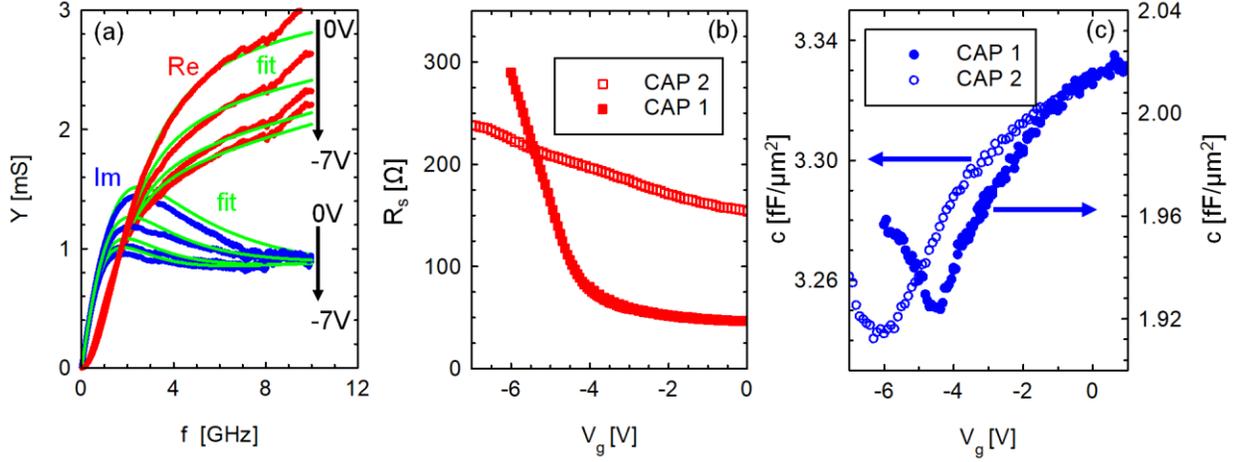

**FIG 3**. Real (red) and imaginary (blue) parts of the admittance of CAP2 versus frequency up to 10GHz at different gate voltage between 0V and -7V. Green curves are theoretical fits using the distributed RC model. (b) Square resistance and (c) capacitance versus gate voltage at 10K. In (b) and (c) filled and open symbols denote the parameters of CAP1 and CAP2 respectively.

The real and imaginary parts of Y obtained from CAP2 are plotted in Fig. 3(a). Those of CAP1 are shown in our previous work. [15] The admittance agrees well with that expected for a 1D distributed RC model [1] with a series contact resistance (Fig. 2(e)) $R_{contact}$:

$$Y_{total}(\omega) = \frac{1}{Y_{RC}^{-1}(\omega) + R_{contact}}, \quad (2)$$

Where,

$$Y_{RC}(\omega) = j\omega c LW \frac{\tanh(L\sqrt{j\omega c \sigma^{-1}})}{L\sqrt{j\omega R c \sigma^{-1}}}, \quad (3)$$

Here, $\omega$ is the frequency. $\sigma$ is the channel conductance (in $\Omega^{-1}$) and $c$ is the total device capacitance per unit area. By curve fitting the real and imaginary parts of Y we can obtain three fit parameters shown in Eq. (2), $\sigma$, $c$ and the contact resistance $R_{contact}$. The latter is fixed to 150$\Omega$. Allowing it to vary yields variations that are negligible within the uncertainty of the experiment.

The channel square resistance $R_s = (\sigma^{-1})$ and capacitance per unit area $c$ ($= C/LW$) for both devices are shown in Fig. 3(b,c) respectively. $L$ and $W$ are respectively the length and width of the capacitor. While a clear capacitance minimum is observed in both CAP1 and CAP2, suggesting successful tuning through the Dirac node, the channel resistance does not exhibit a maximum, indicating the presence of additional conducting channels in the samples that are not well coupled to the gate. The analysis of the quantum capacitance sheds further light on this issue. The data shown in Fig. 3(c) can be used to extract the quantum capacitance $c_q$. From the capacitance per unit area, we extract $c_q$ using $c = \left(c_g^{-1} + c_q^{-1}\right)^{-1}$. The geometric capacitance $c_g$ for both devices is shown in table 1. Notice that $c_g$ is smaller in CAP1 since the gated surface is exposed to air before the hBN is transferred in top. This introduces impurities – possibly an air or vacuum layer between the hBN and $Bi_2Se_3$ – and reduces the dielectric constant of the gate dielectric. In CAP2, $c_g$ yields the dielectric constant of hBN that is recognized in the literature ($\kappa \approx 3.2$). [7] [21] [22] The data in Fig. 3(c) also allows us to extract the chemical potential at the gated surface using the Berglund integral [6]:

$$\Delta\varepsilon_f = \int_0^{V_g} dV \left(1 - \frac{C(V)}{c_g}\right)$$

We are thus able to extract how the quantum capacitance – a direct measurement of the density-of-states – varies with the surface chemical potential ($\Delta\varepsilon_f$) relative to its position at zero gate voltage. This is shown in Fig. 4(a). A linear fit is performed in Fig. 4(a) to determine the Dirac velocity using Eq. (1). We find $5.2\times10^5$m/s and $4.9\times10^5$m/s for CAP1 and CAP2 respectively in agreement with previous studies. [23] [24] The velocity of Dirac particles is thus identical for a TI surface and the TI/hBN interface within our experimental uncertainty.

An important feature of the Fig. 4(a) is the finite residual density-of-states that is reproducibly observed in both samples. It is easy to rule out calibration imperfections and random experimental error as the origin of this finite density of states at zero chemical potential. While the quantum capacitance of ideal Dirac cones is never zero at the Dirac node, previous measurements on graphene have observed order of magnitude lower values. [25] Additionally, even if the quantitative value of $c_{min}$ is highly sensitive to the value of $c_g$, $c_{min}$ remains significant within the experimental error bars associated with $c_g$ (see Table 1). It is thus obvious that $c_{min}$ is not an experimental artifact.

Since the bottom surface is not efficiently coupled to the gate, our devices always result in a chemical potential offset between the two surfaces. We have previously analyzed the capacitance minimum in $Bi_2Se_3$ (in CAP1) and have attributed it to the capacitive coupling through the depleted bulk of the ungated surface with the gated surface due to the large permittivity of Bi2Se3. [15] When the chemical potential is at the Dirac point at the top surface and the bulk is depleted, the ungated bottom surface is capacitively coupled through the insulating bulk. In CAP1, this yields:

$$c_{min} = \left(\frac{1}{c_g^{Bulk}} + \frac{1}{c_q^{ungated}}\right)^{-1} \approx 30 fF/\mu m^2$$

With,

$$c_g^{Bulk} = \frac{\kappa\varepsilon_0}{d_{Bi2Se3}} = 110\ fF/\mu m^2$$

Here the dielectric constant $\kappa = 100$ [26] [27] [28] [29] and $d_{Bi2Se3} = 8nm$. $c_q^{ungated} \approx 40 fF/\mu m^2$ Using Eq. 1 we can show that this corresponds to a Fermi energy of about 170meV above the Dirac point at the bottom ungated surface.

It is interesting to note that the offset observed in CAP2 with the top surface exposed to air, is larger than the one observed in CAP1 with the passivated surface, even within experimental uncertainty. We find $c_{min} > 50 fF/\mu m^2$ for the upper bound of the uncertainty in $c_g$ (see table 1). If we assume an ungated surface coupled to the gated surface of CAP2 through an insulating bulk with we can compute $c_{min}$ using:

$$c_{min} = \left(\frac{1}{c_g^{Bulk}} + \frac{1}{c_q^{ungated}}\right)^{-1} > 50 fF/\mu m^2$$

With $c_g^{Bulk} = 110 fF/\mu m^2$, one needs $c_q^{ungated} > 87 fF/\mu m^2$. From Eq. 1, this corresponds to a Fermi energy in excess of 390meV above the Dirac point if only Dirac fermions are assumed to populate the ungated surface. In Bi$_2$Se$_3$, the Dirac point is about 200meV below the conduction band, [23] [24] therefore a conduction level has to be populated at the ungated surface. We thus have to consider the presence of a surface accumulation layer to account for $c_{min}$ in CAP2. The quantum capacitance from this layer is given by,

$$c_q^{2DEG} = \frac{e^2 m}{\pi \hbar^2} = 94 \, fF/\mu m^2$$

Here $m = 0.14 m_0$ is the effective mass of the 2DEG taken equal to that of the bulk conduction band of Bi$_2$Se$_3$. [30] With $c_q^{ungated} = c_q^{2DEG}$ determined above, $c_{min}$ can be greater than $50 fF/\mu m^2$. Note that it is obviously possible to have contribution from both a Dirac surface states and a 2DEG yielding $c_q^{ungated} = c_q^{2DEG} + c_q^{TSS}$. However, it is not straightforward to determine each independently given the uncertainty on $c_{min}$.

The parameters extracted from the experiment for CAP1 and CAP2 are summarized in table 1. From the analysis of Fig. 4(a), we thus conclude that in CAP1, when the top gated surface state is charge neutral and the bulk is depleted, the bottom ungated topological interface state is populated with a Fermi energy about 170meV above its Dirac point (Fig. 4(b)). In the case of CAP2, when the bottom gated interface state is charge neutral, the bare top surface is also populated by an electron accumulation layer (Fig. 4(c)). Without this accumulation layer, the finite capacitance minimum cannot be accounted for in CAP2.

Differences in sample quality resulting from growth might also explain $c_{min}$ being larger for CAP2. Results from three different growths were, however, analyzed in our previous work and the quantum capacitance in the trilayers hBN-Bi$_2$Se$_3$-hBN consistently reached values smaller than $35 fF/\mu m^2$. [15] All these observations suggest that the presence of the bare surface exposed to atmosphere likely results in the accumulation layer responsible for the additional capacitance observed in CAP2. Note that surface and interface roughness can also yield a finite capacitance by causing a spatial inhomogeneity of the position of the Dirac point in energy. However, this would require the disorder potential close to 100meV to account for the observed $c_{min}$, and would not entirely explain the differences observed by comparing devices. [31]

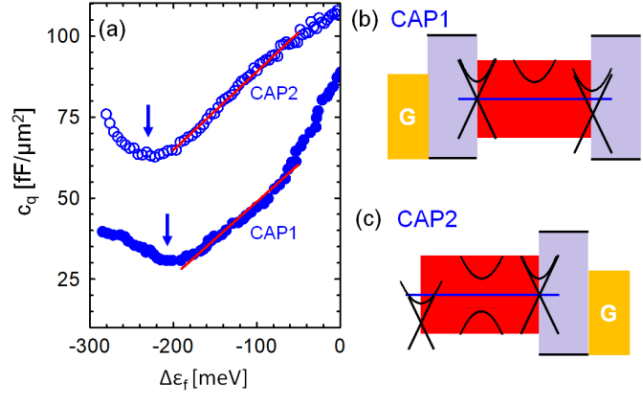

**FIG 4**. (a) Quantum capacitance versus Fermi energy for CAP1 (full circles) and CAP2 (empty circles). Red lines are curve fits using Eq. (1). Band alignment diagram of CAP1 (b) and CAP2 (c) with the ungated surface offset in energy with respect to the gated surface. G marks the location of the gate.

The reason why the resistance (Fig. 3(b)) does not reach a maximum at the quantum capacitance minimum in both CAP1 and CAP2 is also likely due to these multiple contributions from the ungated surfaces. In CAP1, the top and bottom surface are both conducting, but the bottom surface retains a large number of carriers when the top surface is neutral. The same situation arises in CAP2, but the presence of the trivial accumulation layer likely yields a weaker the modulation of the resistance versus gate voltage as seen in Fig. 2(b). This observation can be qualitatively explained by the fact that a parabolic dispersing 2DEG has a flat density of states (versus Fermi energy). However, a quantitative understanding of the behavior of the resistivity requires further studies.

Lastly, note that previous photoemission and transport measurements [27] [32] have reported the observation of a surface 2DEG coexisting with the topological state. Additional measurements also showed that the Dirac points at opposite surfaces can be offset in energy by several 100meV. [33] Our results agree with those observations and suggest that this 2DEG can have a parasitic impact on Dirac transport if left un-passivated or un-gated. Even under a strong electric field (0.75V/nm) applied to one surface of a TI and likely strong enough to deplete the bulk bands, an accumulation layer at the opposite ungated surface can remain difficult to alleviate if that surface has been exposed to air. The hBN layer likely prevents the formation of the accumulation layer by protecting the ungated surface during the device processing steps. Interestingly, even though the conductivity of the TI remains dominated by transport from carriers at the ungated surface and interfacial layer, the quantum capacitance of the Dirac cones at the gated surface can still be reliably measured.

| Device | CAP1 | CAP2 |
|---|---|---|
| Gate configuration | Top gate | Bottom gate |
| Structure | hBN-$Bi_2Se_3$-hBN-Gate | Gate-hBN-$Bi_2Se_3$ |
| LxW [μm x μm] | 4.3 × 12.8 | 20 × 3 |
| hBN thickness | 8.0nm (12 bilayers) | 8.3±0.2nm (12-13 bilayers) |
| $Bi_2Se_3$ thickness (±1QL) | 8QL | 8QL |
| $c_g$ [fF/μm$^2$] | 2.25±0.02 | 3.41±0.05 |
| $\Delta\varepsilon_f$ (at gated Dirac point) | 195meV | 220meV |
| $v_D$ | 5.2×10$^5$ m/s | 4.9×10$^5$ m/s |
| $c_g^{Bulk}$ [fF/μm$^2$] | 110 | 110 |
| $c_{min}$ [fF/μm$^2$] | 25 to 35 | 50 to 85 |

Table 1. Device parameters of CAP1 and CAP2.

In summary, we have measured the RF quantum capacitance and channel resistance of two MITI-CAP devices. The quantum capacitance shows a behavior reminiscent of Dirac fermions characteristic of the surfaces of Bi$_2$Se$_3$. Our measurements demonstrate the importance of passivating and gating away charges from surfaces exposed to air to achieve surface conduction, in the addition to optimizing the bulk of Bi$_2$Se$_3$ to become insulating. Several recent developments have yielded promising results in this direction. [34] [35] [16] Most importantly, our work shows that the RF compressibility remains a viable technique to study the fundamental dielectric properties of topological surfaces in the presence of a finite and small parasitic density-of-states from other bands. As long as a gate voltage can be reliably used to locally tune the Fermi energy across the Dirac point, the compressibility of the Dirac state can be measured. This fact becomes particularly important when considering metallic systems such as three-dimensional Weyl semimetals with Fermi arcs at their surfaces. [36] [37]

**Acknowledgements.** We acknowledge Michael Rosticher for valuable assistance with cleanroom equipment and Pascal Morfin for his assistance with the growth setup. We also acknowledge discussions with H. Graef. BAA, JD and TW acknowledge funding from ANR-LabEx grant ENS-ICFP ANR-10-LABX-0010/ANR-10-IDEX-0001-02PSL. JD was partially supported by the Northeastern University Coop fellowship. K.W. and T.T. acknowledge support from the Elemental Strategy Initiative conducted by the MEXT, Japan, A3 Foresight by JSPS and the CREST (JPMJCR15F3), JST.

[1] E. Pallecchi, a. C. Betz, J. Chaste, G. Fève, B. Huard, T. Kontos, J.-M. Berroir, and B. Plaçais, "Transport scattering time probed through rf admittance of a graphene capacitor". Phys. Rev. B. **83**, 125408 (2011).

[2] A. C. Betz, S. H. Jhang, E. Pallecchi, R. Ferreira, G. Fève, J.-M. Berroir, and B. Plaçais, "Supercollision cooling in undoped graphene". Nat. Phys. **9**, 109 (2012).

[3] J. Chaste, L. Lechner, P. Morfin, G. Fève, T. Kontos, J. M. Berroir, D. C. Glattli, H. Happy, P. Hakonen, and B. Plaçais, "Single carbon nanotube transistor at GHz frequency". Nano Lett. **8**, 525 (2008).

[4] G. Fève, A. Mahé, J.-M. Berroir, T. Kontos, B. Plaçais, D. C. Glattli, A. Cavanna, B. Etienne, Y. Jin, G. Feve, A. Mahe, J.-M. Berroir, T. Kontos, B. Placais, D. C. Glattli, A. Cavanna, B. Etienne, and Y. Jin, "An On-Demand Coherent Single-Electron Source". Science (80-. ). **316**, 1169 (2007).

[5] Y. Xu, C. Chen, V. V. Deshpande, F. A. DiRenno, A. Gondarenko, D. B. Heinz, S. Liu, P. Kim, and J. Hone, "Radio frequency electrical transduction of graphene mechanical resonators". Appl. Phys. Lett. **97**, 243111 (2010).

[6] C. N. Berglund, "Surface states at steam-grown silicon-silicon dioxide interfaces". IEEE Trans. Electron Devices. **ED-13**, 701 (1966).

[7] H. Graef, D. Mele, M. Rosticher, L. Banszerus, C. Stampfer, T. Taniguchi, K. Watanabe, E. Bocquillon, G. Fève, J. Berroir, E. H. T. Teo, and B. Plaçais, "Ultra-long wavelength Dirac plasmons in graphene capacitors". J. Phys. Mater. **1**, 01LT02 (2018).

[8] W. Yang, S. Berthou, X. Lu, Q. Wilmart, A. Denis, M. Rosticher, T. Taniguchi, K. Watanabe, G. Fève, J. Berroir, G. Zhang, C. Voisin, E. Baudin, and B. Plaçais, "A graphene Zener–Klein

transistor cooled by a hyperbolic substrate". Nat. Nanotechnol. **13**, 47 (2018).

[9] A. Akhmerov, J. Nilsson, and C. Beenakker, "Electrically Detected Interferometry of Majorana Fermions in a Topological Insulator". Phys. Rev. Lett. **102**, 216404 (2009).

[10] R. A. Snyder, C. J. Trimble, C. C. Rong, P. A. Folkes, P. J. Taylor, and J. R. Williams, "Weak-link Josephson Junctions Made from Topological Crystalline Insulators". Phys. Rev. Lett. **121**, 097701 (2018).

[11] R. Klett, J. Schönle, A. Becker, D. Dyck, K. Borisov, K. Rott, D. Ramermann, B. Büker, J. Haskenhoff, J. Krieft, T. Hübner, O. Reimer, C. Shekhar, J. Schmalhorst, A. Hütten, C. Felser, W. Wernsdorfer, and G. Reiss, "Proximity-Induced Superconductivity and Quantum Interference in Topological Crystalline Insulator SnTe Thin-Film Devices". Nano Lett. **18**, 1264 (2018).

[12] E. Bocquillon, R. S. Deacon, J. Wiedenmann, P. Leubner, T. M. Klapwijk, C. Brüne, K. Ishibashi, H. Buhmann, and L. W. Molenkamp, "Gapless Andreev bound states in the quantum spin Hall insulator HgTe". Nat. Nanotechnol. **12**, 137 (2017).

[13] M. C. Dartiailh, S. Hartinger, A. Gourmelon, K. Bendias, H. Bartolomei, J.-M. Berroir, G. Fève, B. Plaçais, L. Lunczer, R. Schlereth, H. Buhmann, L. W. Molenkamp, and E. Bocquillon, "Dressed topological edge states in HgTe-based 2D topological insulators". Arxiv. 1903.12391 (2019).

[14] A. Inhofer, S. Tchoumakov, B. A. Assaf, G. Fève, J. M. M. Berroir, V. Jouffrey, D. Carpentier, M. O. O. Goerbig, B. Plaçais, K. Bendias, D. M. M. Mahler, E. Bocquillon, R. Schlereth, C. Brüne, H. Buhmann, and L. W. W. Molenkamp, "Observation of Volkov-Pankratov states in topological HgTe heterojunctions using high-frequency compressibility". Phys. Rev. B. **96**, 195104 (2017).

[15] A. Inhofer, J. Duffy, M. Boukhicha, E. Bocquillon, J. Palomo, K. Watanabe, T. Taniguchi, I. Estève, J. M. Berroir, G. Fève, B. Plaçais, and B. A. Assaf, "rf Quantum Capacitance of the Topological Insulator $Bi_2Se_3$ in the Bulk Depleted Regime for Field-Effect Transistors". Phys. Rev. Appl. **9**, 024022 (2018).

[16] S. Xu, Y. Han, X. Chen, Z. Wu, L. Wang, T. Han, W. Ye, H. Lu, G. Long, Y. Wu, J. Lin, Y. Cai, K. M. Ho, Y. He, and N. Wang, "van der Waals Epitaxial Growth of Atomically Thin $Bi_2Se_3$ and Thickness-Dependent Topological Phase Transition". Nano Lett. **15**, 2645 (2015).

[17] A. S. Mayorov, R. V Gorbachev, S. V Morozov, L. Britnell, R. Jalil, L. a Ponomarenko, P. Blake, K. S. Novoselov, K. Watanabe, T. Taniguchi, and a. K. Geim, "Micrometer-Scale Ballistic Transport in Encapsulated Graphene at Room Temperature". Nano Lett. **11**, 2396 (2011).

[18] M. H. D. Guimarães, P. J. Zomer, J. Ingla-Aynés, J. C. Brant, N. Tombros, and B. J. van Wees, "Controlling Spin Relaxation in Hexagonal BN-Encapsulated Graphene with a Transverse Electric Field". Phys. Rev. Lett. **113**, 086602 (2014).

[19] W. Richter, H. Köhler, and C. R. Becker, "A Raman and Far-Infrared Investigation of Phonons in the Rhombohedral V2-VI3 Compounds". Phys. Status Solidi. **84**, 619 (1977).

[20] D. M. Pozar, *Microwave Engineering*, 3rd ed. (Wiley, 2005).

[21] K. K. Kim, A. Hsu, X. Jia, S. M. Kim, Y. Shi, M. Dresselhaus, T. Palacios, and J. Kong, "Synthesis and Characterization of Hexagonal Boron Nitride Film as a Dielectric Layer for Graphene Devices". ACS Nano. **6**, 8583 (2012).

[22] C. R. Dean, a F. Young, I. Meric, C. Lee, L. Wang, S. Sorgenfrei, K. Watanabe, T. Taniguchi, P. Kim, K. L. Shepard, and J. Hone, "Boron nitride substrates for high-quality graphene electronics.".

Nat. Nanotechnol. **5**, 722 (2010).

[23] Y. Xia, D. Qian, D. Hsieh, L. Wray, A. Pal, H. Lin, A. Bansil, D. Grauer, Y. S. Hor, R. J. Cava, and M. Z. Hasan, "Observation of a large-gap topological-insulator class with a single Dirac cone on the surface". Nat Phys. **5**, 398 (2009).

[24] J. G. Analytis, J.-H. Chu, Y. Chen, F. Corredor, R. D. McDonald, Z. X. Shen, and I. R. Fisher, "Bulk Fermi surface coexistence with Dirac surface state in Bi2Se3 : A comparison of photoemission". Phys. Rev. B. **81**, 205407 (2010).

[25] G. L. Yu, R. Jalil, B. Belle, A. S. Mayorov, P. Blake, F. Schedin, S. V Morozov, L. a Ponomarenko, F. Chiappini, S. Wiedmann, U. Zeitler, M. I. Katsnelson, a K. Geim, K. S. Novoselov, and D. C. Elias, "Interaction phenomena in graphene seen through quantum capacitance.". Proc. Natl. Acad. Sci. U. S. A. **110**, 3282 (2013).

[26] D. Kim, S. Cho, N. P. Butch, P. Syers, K. Kirshenbaum, S. Adam, J. Paglione, and M. S. Fuhrer, "Surface conduction of topological Dirac electrons in bulk insulating Bi2Se3". Nat. Phys. **8**, 460 (2012).

[27] M. Bianchi, D. Guan, S. Bao, J. Mi, B. B. Iversen, P. D. C. King, and P. Hofmann, "Coexistence of the topological state and a two-dimensional electron gas on the surface of Bi(2)Se(3).". Nat. Commun. **1**, 128 (2010).

[28] N. P. Butch, K. Kirshenbaum, P. Syers, A. B. Sushkov, G. S. Jenkins, H. D. Drew, and J. Paglione, "Strong surface scattering in ultrahigh-mobility Bi2Se3 topological insulator crystals". Phys. Rev. B. **81**, 241301 (2010).

[29] H. Köhler and C. R. Becker, "Optically Active Lattice Vibrations in Bi2Se3". Phys. Status Solidi. **61**, 533 (1974).

[30] M. Orlita, B. a. Piot, G. Martinez, N. K. S. Kumar, C. Faugeras, M. Potemski, C. Michel, E. M. Hankiewicz, T. Brauner, Č. Drašar, S. Schreyeck, S. Grauer, K. Brunner, C. Gould, C. Brüne, and L. W. Molenkamp, "Magneto-Optics of Massive Dirac Fermions in Bulk Bi2Se3". Phys. Rev. Lett. **114**, 186401 (2015).

[31] C. Parra, T. H. Rodrigues da Cunha, A. W. Contryman, D. Kong, F. Montero-Silva, P. H. Rezende Gonçalves, D. D. Dos Reis, P. Giraldo-Gallo, R. Segura, F. Olivares, F. Niestemski, Y. Cui, R. Magalhaes-Paniago, and H. C. Manoharan, "Phase Separation of Dirac Electrons in Topological Insulators at the Spatial Limit". Nano Lett. **17**, 97 (2017).

[32] J. Suh, D. Fu, X. Liu, J. K. Furdyna, K. M. Yu, W. Walukiewicz, and J. Wu, "Fermi-level stabilization in the topological insulators Bi2Se3 and Bi2Te3: Origin of the surface electron gas". Phys. Rev. B. **89**, 115307 (2014).

[33] M. H. Berntsen, O. Götberg, B. M. Wojek, and O. Tjernberg, "Direct observation of decoupled Dirac states at the interface between topological and normal insulators". Phys. Rev. B. **88**, 195132 (2013).

[34] N. Koirala, M. Brahlek, M. Salehi, L. Wu, J. Dai, J. Waugh, T. Nummy, M. G. Han, J. Moon, Y. Zhu, D. Dessau, W. Wu, N. P. Armitage, and S. Oh, "Record Surface State Mobility and Quantum Hall Effect in Topological Insulator Thin Films via Interface Engineering". Nano Lett. **15**, 8245 (2015).

[35] J. Y. Park, G.-H. Lee, J. Jo, A. K. Cheng, H. Yoon, K. Watanabe, T. Taniguchi, M. Kim, P. Kim, and G.-C. Yi, "Molecular beam epitaxial growth and electronic transport properties of high quality topological insulator Bi 2 Se 3 thin films on hexagonal boron nitride". 2D Mater. **3**, 035029

(2016).

[36] S.-Y. Xu, I. Belopolski, N. Alidoust, M. Neupane, G. Bian, C. Zhang, R. Sankar, G. Chang, Z. Yuan, C.-C. Lee, S.-M. Huang, H. Zheng, J. Ma, D. S. Sanchez, B. Wang, A. Bansil, F. Chou, P. P. Shibayev, H. Lin, S. Jia, and M. Z. Hasan, "Discovery of a Weyl Fermion semimetal and topological Fermi arcs". Science (80-. ). **349**, 613 (2015).

[37] K. Deng, G. Wan, P. Deng, K. Zhang, S. Ding, E. Wang, M. Yan, H. Huang, H. Zhang, Z. Xu, J. Denlinger, A. Fedorov, H. Yang, W. Duan, H. Yao, Y. Wu, S. Fan, H. Zhang, X. Chen, and S. Zhou, "Experimental observation of topological Fermi arcs in type-II Weyl semimetal MoTe2". Nat. Phys. **12**, 1105 (2016).